\documentclass[epj, nopacs]{svjour}

\usepackage[english]{babel}

\usepackage{amsmath}
\usepackage{amsfonts} 
\usepackage{amssymb}
\usepackage{cite}
\usepackage{graphicx}

\usepackage{caption}
\addto\captionsenglish{}

\usepackage{ifthen}

\usepackage{xcolor}

\newcommand{\etal}{\textit{et\,al.} }

\newcommand{\Nr}[1]{\ensuremath{ \ifthenelse{\equal{#1}{}}{ \mathbf{r} }{ \mathbf{r}_#1  } }}
\newcommand{\Np}[1]{\ensuremath{ \ifthenelse{\equal{#1}{}}{ \mathbf{p} }{ \mathbf{p}_#1  } }}
\newcommand{\Nphi}[1]{\ensuremath{ \ifthenelse{\equal{#1}{}}{ \varphi }{ \varphi_#1  } }}
\newcommand{\Nu}[1]{\ensuremath{ \ifthenelse{\equal{#1}{}}{ \mathbf{\hat{u}} }{ \mathbf{\hat{u}}_#1  } }}
\newcommand{\Nfunctional}{\ensuremath{\mathcal{F}}}

\newcommand{\Npart}[1]{\ensuremath{\frac{\partial #1}{\partial t}}}
\newcommand{\Nint}[3]{\ensuremath{\int_{#1}^{#2} \! \! d#3}}
\newcommand{\norm}[1]{\left\lVert #1 \right\rVert}

\begin{document}
\title{Time scales of inertial motion in the collective dynamics of underdamped active phase field crystal systems}
\titlerunning{Inertia time scales in underdamped active phase field crystal systems}

\author{Dominic Arold \and Michael Schmiedeberg
\thanks{\email{michael.schmiedeberg@fau.de}}%
}


\institute{Institut f\"ur Theoretische Physik I, Friedrich-Alexander-Universit\"at Erlangen-N\"urnberg, Staudtstra\ss e 7, 91058 Erlangen, Germany}
\date{Received: date / Revised version: date}
%
\abstract{
Many active matter systems, mostly on the microscopic scale, are well approximated as overdamped, meaning that any inertial momentum is immediately dissipated by the environment. On the other hand, especially for macroscopic active systems but also for many mesoscopic ones the time scale of inertial motion can become large enough to be relevant for the dynamics. This raises the question how collective dynamics in active matter is influenced by inertia. In this article we implement and study an underdamped active phase field crystal model. We focus on how the collective dynamics changes with the time scale of inertial motion. While the state diagram stays unaltered in this modification, the relaxation time scale towards the steady state considerably increases with particle mass. Our numerical results suggest that transiently stable rotating clusters of density peaks act as defects which need to decay before the final state of global collective motion forms. We extract the formation and decay times quantitatively. Finally, we give a physical intuition for the formation and decay of rotating clusters to qualitatively explain how the extracted times depend on mass.
}

\maketitle

\section{Introduction}
\label{sec:intro}

Many-body systems composed of self-propelled, interacting particles are widely studied in many variations (for reviews, see \cite{Marchetti2013, bechinger2016active,doostmohammadi2018active,Lowen2019Inertial}). Physically, the particles correspond to e.g. biological or artificial microswimmers \cite{wensink2012meso, Heidenreich2016, Menzel2016, Reinken2018a}, cell colonies \cite{Dombrowski2004, riedel2005self, Rossen2014}, or protein filaments \cite{sumino2012large}. In all systems mentioned so far inertia effects usually can be neglected due to the low Reynolds numbers such that viscous forces are much larger than the contributions of inertia. However, there are also systems like flocks of birds \cite{ballerini2008interaction, ginelli2010relevance} or artificial robots \cite{Kudrolli2008, Bricard2015, Deseigne2012} where inertia cannot be neglected. Therefore, many recent works study the consequences of inertial effects in active systems \cite{scholz2018inertial,mandal2019motility,arold2019mean,Lowen2019Inertial}.

A mean field approach to active systems can be obtained by extending the phase field crystal (PFC) model \cite{elder2002modeling, elder2004modeling, elder2007phase} to active systems. This can be achieved by describing the orientation of the active particles with an additional vector field \cite{Menzel2013, Menzel2014}. The peaks in the density field may still be identified as single particles but other interpretations are reasonable as e.g. periodic patterns of accumulated active particles are predicted to form which are comparable to the density peaks in the PFC approach by Menzel \etal\cite{Menzel2013a}. For low activity the peaks order hexagonally while remaining at rest (resting crystal state). However, above a critical active drive peaks start to move due to self-propulsion. A local alignment interaction of orientations induces the formation of translationally migrating clusters of density peaks that coarsen over time. In the long time limit global collective motion is observed (traveling crystal state). Such or similar continuum approaches have been used to describe collective dynamics in bacterial colonies \cite{wensink2012meso}, in ensembles of microswimmers \cite{Heidenreich2016, Ariel2018}, or in active nematics \cite{chandragiri2019active, mueller2019emergence, doostmohammadi2018active}.

In this article we want to extend the PFC approach for active systems such that inertia contributions can be studied as well. Inertia is introduced in a similar way as in our recent work \cite{arold2019mean}. Note that in this approach the direction of the actual velocity and therefore of the inertial effects might be different from the direction of the intrinsic self-propulsion. Here, we focus on the impact that the introduced time scale of inertial motion on the single particle level has on the large scale collective dynamics in active PFC systems.

The article is structured as follows: In section \ref{sec:model} we outline our model for underdamped active matter, use it to extend the overdamped active PFC model, and describe how we analyze our simulation results. The latter are then discussed in section \ref{sec:results} and concluded in section \ref{sec:conclusion}.

\section{Model system and methods}
\label{sec:model}

We use a mean field approach to describe an ensemble of massive and interacting particles which self-propel along their individual polar orientation with a constant force $f_0$. Interaction forces as well as interaction torques between orientations are formulated via free energy functionals $\Nfunctional$ and $\Nfunctional_P$ respectively. A particle might also change its direction of self-propulsion independent of other particles' orientations by applying a one-particle torque $G^{(1)}$. Further, dissipative coupling to an environment is considered as well as translational and rotational diffusion. Altogether, our dynamics of the locally averaged fields for number density $\rho(\Nr{}, t)$, velocity $\mathbf{v}(\Nr{}, t)$ and polar orientational order $\mathbf{P}(\Nr{}, t)$ reads

\begin{equation}
\begin{aligned}
\Npart{\rho} &= - \nabla \cdot (\rho  \textbf{v})\\
\frac{\mathrm{D} \textbf{v}}{\mathrm{D} t} &= \frac{1}{m} \left(-\alpha \, \textbf{v} - \nabla \frac{\delta \Nfunctional}{\delta \rho} + \alpha v_0 \, \textbf{P} \right)\\
\frac{\mathrm{D} \textbf{P}}{\mathrm{D} t} &= -D_R \textbf{P} - \frac{\delta \Nfunctional_P}{\delta \textbf{P}} + \overline{\textbf{G}^{(1)}}
\end{aligned}
\label{eq:model_for_ud_active_systems}
\end{equation}

with the convective derivative $\frac{\mathrm{D} }{\mathrm{D} t} = \frac{\partial}{\partial t} + (\textbf{v} \cdot \nabla)$. The self-propulsion velocity $v_0 = f_0 / \alpha$ corresponds to the steady state velocity of a particle in an environment with friction constant $\alpha$ and accelerated by the active force $f_0$. $D_R$ is the rotational diffusion coefficient and $G^{(1)}$ is locally averaged over the distribution of particle momenta and orientation angles. We emphasize that including inertia into the dynamics introduces an additional time scale given by the inverse damping rate $\gamma^{-1} = m / \alpha$ with particle mass $m$. For our derivation of eqs.~\eqref{eq:model_for_ud_active_systems} via Dynamical Density Functional Theory we refer to \cite{arold2019mean}.

In \cite{arold2019mean} we have considered a system without any preferred length scale in the density field, but an alignment interaction for short distances and an anti-alignment interaction at longer distances. Here we want to explore the properties of an underdamped active PFC model, i.e. with no preferred length scale concerning the alignment interactions, but usual PFC interactions for the density field. To be specific, we choose the unspecified terms in eqs.~\eqref{eq:model_for_ud_active_systems} such that in the limit $m \to 0$ the original overdamped active PFC model by Menzel \etal \cite{Menzel2013, Menzel2014} is recovered. Thus, we expect that results of the latter are reproduced within this generalized underdamped model for small mass. Consequently, the functionals and the mean one body torque are set to 
\begin{equation}
\begin{aligned}
\Nfunctional[\rho] &= \Nint{}{}{\Nr{}} \ \frac{\rho}{2} \left[ \epsilon + \lambda \left(q_0^2 + \nabla^2 \right)^2 \right] \rho + \frac{u}{4} \, \rho^4\\
\Nfunctional_{P}[\mathbf{P}] &= \frac{C_1}{2} \Nint{}{}{\Nr{}} \ \left(\left|\nabla P_x \right|^2 + \left|\nabla P_y \right|^2 \right)\\
\overline{\mathbf{G}^{(1)}} &= - \dfrac{v_0}{\left|\overline{\rho}\right|} \, \nabla \rho.
\end{aligned}
\label{eq:functionals}
\end{equation}

Translational interactions in $\Nfunctional$ are given by the PFC functional \cite{elder2002modeling, elder2004modeling}. We observe that the orientational interaction functional $\Nfunctional_P$ is conceptually equivalent to the Frank-Oseen free energy of liquid crystals with one bending constant $C_1$ \cite{frank1958liquid}. For $C_1 > 0$ a homogeneous state of uniformly aligned particle orientations minimizes the free energy. Contrarily, the one particle torque $\overline{\mathbf{G}^{(1)}}$, whose strength is given by the activity parameter $v_0$, drives an instability from the isotropic state $\mathbf{P} = \mathbf{0}$.
The PFC functional only describes variations of a density around its constant bulk mean value. The parameter field $\rho$ may be interpreted as such after rescaling which justifies a constant mobility approximation for the continuity equation in eqs.~\eqref{eq:model_for_ud_active_systems}. We want to focus in the following on the influence of the inertial time scale $\gamma^{-1}$ on the system's dynamics. To this end we neglect the convective term $(\textbf{v} \cdot \nabla) \,  \textbf{P}$ for now as it qualitatively changes the observable states such that a direct comparison to the overdamped active PFC model is not possible. We will briefly outline the case without this approximation in sect.~\ref{sec:conclusion}. Altogether, the time evolution of the mean fields in the underdamped active PFC model then reads

\begin{equation}
\begin{aligned}
\Npart{\rho} &= - \left| \overline{\rho} \right| \nabla \cdot \textbf{v}\\
\Npart{\textbf{v}} + (\textbf{v} \cdot \nabla) \, \textbf{v} &= \frac{1}{m} \left(-\alpha \, \textbf{v} - \nabla \frac{\delta \Nfunctional}{\delta \rho} + \alpha v_0 \, \textbf{P} \right)\\
\Npart{\textbf{P}}  &= -D_R \, \textbf{P} - \frac{\delta \Nfunctional_P}{\delta \textbf{P}} - \frac{v_0}{\left| \overline{\rho} \right|} \, \nabla \rho.
\end{aligned}
\label{eq:model_ud_crystal}
\end{equation}

By taking the overdamped limit $m \to 0$ the convective derivative on the left hand side of the velocity equation becomes negligible leading to the simplified form 

\begin{equation}
\begin{aligned}
\Npart{\rho} &= \left| \overline{\rho} \right| \left( \dfrac{1}{\alpha} \, \nabla^2 \frac{\delta \Nfunctional}{\delta \rho} - v_0 \nabla \cdot \textbf{P}\ \right)\\
\Npart{\textbf{P}}&= -D_R \, \textbf{P} - \frac{\delta \Nfunctional_P}{\delta \textbf{P}} - \dfrac{v_0}{\left|\overline{\rho}\right|} \, \nabla \rho.
\end{aligned}
\label{eq:model_od_crystal}
\end{equation}

After rescaling, eqs.~\eqref{eq:model_od_crystal} coincide with the active PFC model by Menzel \etal \cite{Menzel2013, Menzel2014} except for the additional parameter $\left| \overline{\rho} \right|$ due to the higher number of parameters. Concerning the Free Energy functional in eq.~\eqref{eq:functionals}, after choosing the length such that $q_0=1$ and rescaling the density variations and the free energy such that $\lambda=u=1$, only two independent parameters remain, the temperature-like $\varepsilon$ as well as the mean density variation $\overline{\rho}$ (cf. \cite{elder2002modeling}). In the following we use $\varepsilon = -0.98$ and $\overline{\rho} = -0.4$ such that for the usual passive PFC model we are in the hexagonal crystal phase \cite{elder2002modeling}. Concerning the parameters in eqs.~\eqref{eq:model_ud_crystal} and \eqref{eq:model_od_crystal}, the time variable and mass can be rescaled such that $\alpha=1$. Furthermore, we fix $D_R=0.1$ and $C_1=0.2$ and focus on the remaining parameters, namely the active drive $v_0$ and mass $m$ which measures the time scale of inertial motion. All numerical calculations discussed in the following are started from homogeneous initial conditions overlaid by a small noise.

\begin{figure*}[ht]
\centering
 \includegraphics[width=\textwidth]{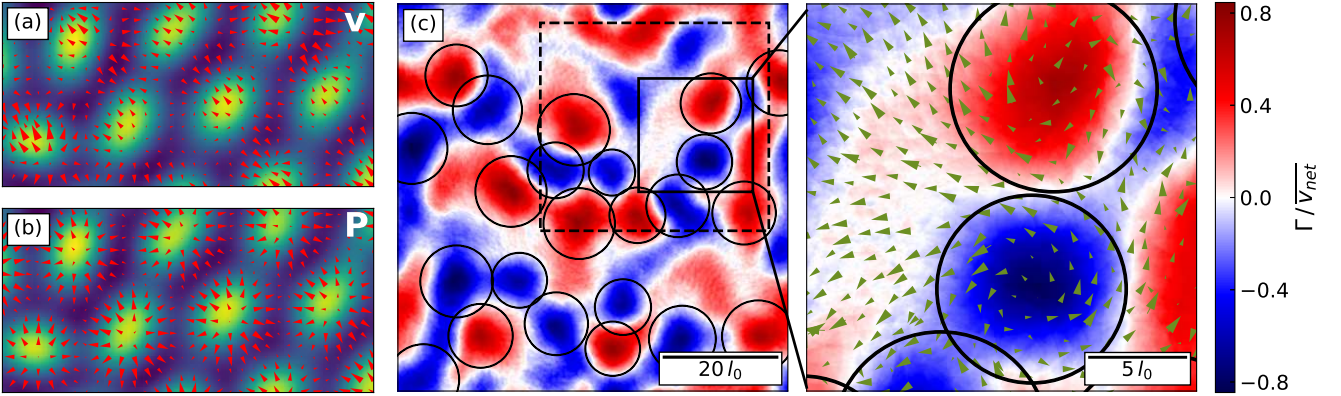}
\caption{Density field during the relaxation to the traveling crystal state with (a) corresponding velocity and (b) orientation field. Positions of density peaks and defects in the orientation field are shifted relative to each over. The resulting net-orientation averaged over a peak area induces self-propulsion of peaks reflected in the non-vanishing velocity field. Arrows in the enlarged view displayed in the right hand side of (c) indicate net velocities $\mathbf{v}_{net}$ of peaks used to compute the shown circulation field $\Gamma$. Rotating clusters of peaks are indicated by circles with radius $R_\Gamma$ centered at extrema of $\Gamma$. A time lapse of the density field in the region indicated by dashed lines is given in the supplemented movie \cite{supplement}, where one can see the rotating clusters that occur during intermediate times. The length scale is shown in units of the preferred PFC length scale $l_0 = 2\pi / q_0$. Parameters are $m=2$ and $v_0=0.4$. The snapshot is taken after a simulation time $t = 500$.}
\label{fig1}
\end{figure*}

To analyze our numerical data we track the position of density peaks and assign a velocity to each. Especially for higher $m$ and $v_0$ values the method in \cite{Menzel2013, Menzel2014} used for this becomes unreliable here since then fluctuations in the density field increase and the peak structure is not as clear as in the overdamped limit (compare supplemented movie \cite{supplement}). Instead, we first locate maxima in the density field and identify to each such found peak the area around it where the density exceeds a threshold. By averaging the velocity field $\textbf{v}$ over this area we assign to each peak a momentary velocity $v_{net}$. As can be seen in fig.\ref{fig1}~(a) the velocity field $\textbf{v}$ at a density peak is maximal close to the peak maximum around which it is also averaged to $v_{net}$. Because of this the latter overestimates the actual peak velocity in the sense of a displacement velocity. But for the following discussion of the results the absolute values of velocities are irrelevant since we only consider relative values. By restricting to area points within a given maximal distance to their corresponding peak position and a minimal distance between peak positions we suppress the influence of fluctuations in the density field.

\section{Results}
\label{sec:results}

Generally we find that the under- and overdamped model given by eqs.~\eqref{eq:model_ud_crystal} and \eqref{eq:model_od_crystal}, respectively, predict the same steady states in the long time limit. Hence, we only briefly recap them here and refer to the results for the overdamped model studied in \cite{Menzel2013, Menzel2014} for more details. For low activity $v_0$ the system is in the resting crystal state. The kinetic energy input due to activity of the particles melts crystals close to the liquid-solid phase boundary and therefore shifts the latter to lower temperatures by $\Delta \epsilon$. The crystalline structures remain at rest. Numerics and linear stability analysis of the resting crystal state within the underdamped model predict $\Delta \epsilon \propto v_0^2 \alpha / C_1$ which is in accordance with the findings of Menzel \etal \cite{Menzel2013}. Above the critical activity $v_{0, c} \approx 0.3$ the phase diagram is not shifted further to lower temperatures. Instead, the additional kinetic energy input from activity is used for translational self-propulsion of the density peaks as depicted in fig.~\ref{fig1}~(a) and (b). Under- and overdamped model predict the same traveling crystal state in the long time limit where all hexagonally ordered density peaks have aligned their migration direction and move with the same velocity.

The mass parameter does not change the traveling crystal state in the long time limit for $v_0 > v_{0, c}$. Therefore, introducing the inertial time scale $\gamma^{-1} = m / \alpha$ does not qualitatively change the non-equilibrium state. What indeed changes is the time scale of the relaxation process to the steady state. As we will show in the following, the relaxation considerably slows down with $m$ and therefore cannot be characterized with the inertial time scale $\gamma^{-1}$ that is orders of magnitude smaller. Therefore, additional collective effects arising from particle interactions influence the transient behavior of the system in dependence of the inertial regime.

Typical peak velocity patterns, like the ones shown in fig.~\ref{fig1}~(c), and the supplemented time lapse of the corresponding density field \cite{supplement} suggest that during transient time scales the system is not only characterized by translationally moving clusters of density peaks as observed in the overdamped model \cite{Menzel2013} but also by rotational ones. Characterizing the latter will help to quantify and explain the relaxation time scale.

\begin{figure}[t]
\centering
 \includegraphics[width=\columnwidth]{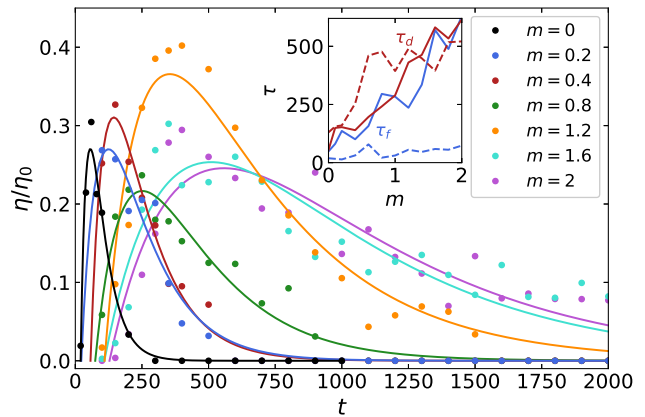}
\caption{The time courses of the area fraction $\eta$ for different mass values can be understood in terms of the phenomenological expression in eq.~\eqref{eq:eta_fit} (solid lines). Inset: Extracted formation times $\tau_f$ (blue) and decay times $\tau_d$ (red) systematically depend on particle mass. Solid lines are results for $v_0 = 0.35$ while dashed lines correspond to the case $v_0 = 0.4$. $m=0$ results are obtained from the overdamped limit dynamics in eqs.~\eqref{eq:model_od_crystal}.}
\label{fig2}
\end{figure}

To this end we use the concept of circulation which we define for a continuous velocity field $\textbf{v}$ as the closed line integral $\Gamma_c(r) = \oint_{\partial A} \mathbf{v} \cdot d\mathbf{l}$ over the boundary of a surface $A$. Via Stokes theorem this can also be understood as the flux of the vorticity field $\mathbf{\omega} = \nabla \times \mathbf{v}$ through $A$ and thus measures the amount of circular motion in this area. Similarly, we compute a circulation for the discrete peak velocities $v_{net}$ at any given point $\mathbf{r}$ by averaging the velocity components in angular direction $\hat{\mathbf{e}}_l$ around $\mathbf{r}$ over all peaks within a radius $R_{\mathrm{max}}$ around this point. Formally, the circulation then reads
\begin{equation}
\Gamma(\mathbf{r}) = \overline{\mathbf{v}_{net}(\mathbf{r}+\mathbf{R}) \cdot \hat{\mathbf{e}}_l}
\label{eq:circulation}
\end{equation}
where the average runs over all relative peak positions with $\norm{\mathbf{R}} < R_{\mathrm{max}}$. As can be seen seen in the magnified region of fig.~\ref{fig1}~(c) the circulation has its extrema at the centers of rotating clusters. For each sample we count the number $N_\Gamma$ of extrema as a function of time. To better distinguish rotational from translational moving clusters the condition $\left| \Gamma(\mathbf{r}) \right| > \tfrac{2}{\pi} \overline{v_{net}}(t)$ for possible extrema is used. This threshold is the maximum circulation at the boundary between two anti-parallel moving translational clusters which move with the momentary sample-averaged peak velocity. We define the size of a rotating cluster as the area around an extremum in $\Gamma$ with $\left| \Gamma(\mathbf{r}) \right| > 0$. In fig.~\ref{fig1}~(c) this area is depicted for each cluster as a circle of equal area with radius $R_\Gamma$. An upper bound for the distance between an area element and the extremum position is used to avoid including adjacent areas of dominant translational motion into the area estimation. The presence of transiently stable rotating clusters delays the relaxation to the final steady state of global collective motion. We extract the corresponding time scales from the temporal evolution of the area fraction of rotating clusters in the simulation box with extensions $l_x$ and $l_y$ which reads
\begin{equation}
\eta = \frac{N_\Gamma \, \pi \overline{R_\Gamma^2}}{l_x l_y}.
\label{eq:filling_factor}
\end{equation}
This number measures how much the momentary global dynamics is governed by local clusters of circular motion. A value of $\eta = \pi / 4$ corresponds to the extreme case of a frustration free square lattice of equally sized clusters where each is surrounded by four contrary rotating ones. The time courses of $\eta$ in fig.~\ref{fig2} generally show an initial rise, reflecting the formation and growth of rotating clusters over a period $\tau_f$, followed by a decline period $\tau_d$ during which clusters break-up again. Afterward, the system approaches the steady traveling crystal state where $N_\Gamma = 0$.

Motivated by these qualitative systematics we suggest a phenomenological expression for the area fraction. Treating $\tau_f$ and $\tau_d$ as exponential time scales leads to the expression
\begin{equation}
\eta(t) = \eta_0 \left( 1- e^{-t/\tau_f} \right) e^{-t/\tau_d}.
\label{eq:eta_fit}
\end{equation}
We extract the mass and activity dependent formation and decay time scales by fitting this function to the data in fig.~\ref{fig2}. The value $\eta_0 = \pi / 4$ stays fixed and a variable time offset is used to account for the varying onsets of crystallization from the supercooled liquid.

The observed transient time scales are shown in fig.~\ref{fig2}. They represent the main result of this article as they summarize the impact of single particle inertia on the collective dynamics. In order to physically explain their dependence on the parameters we propose an idealized conception for the formation and break-up mechanism of rotating clusters.

After the initialization from isotropic initial conditions smaller translationally moving density peak clusters form. At the boundary between two contrarily moving ones, peaks may switch from one of these common migration directions to the other by rotating their orientation of self-propulsion. If peaks from both clusters are doing so they align with each other too while moving further past each other thereby having to rotate their orientations even further. This disturbance of the translational cluster boundary eventually drives the emergence of a rotating cluster which grows as more peaks align to it. When a peak rotates its orientation, the change in motion happens instantly in the overdamped limit. On the other hand, when inertia is relevant the actual motion of particles does not follow a change in orientation immediately. Instead of joining the contrarily moving cluster a peak might reorient back to its current one. Therefore, the probability of initializing a rotating cluster decreases with mass and its growth process is delayed to later times. Therefore, the maximum number of rotating clusters is expected to decrease with increasing $m$. Furthermore, $\tau_f$ increases with $m$ as we observe for $v_0 = 0.35$.

We also find that cluster radii saturate when equally many peaks align to and detach from a cluster. Rotating clusters then further persist over the decay time scale $\tau_d$ due to the sufficiently given local alignment. However, after some time the systems' preference for a hexagonal peak structure prevails. Then all rotational defects decay and translational clusters with hexagonal symmetry coarse-grain to the final traveling crystal state. Instead of a steady dissolution, the decay process of rotating clusters is better described as a rather abrupt break-up into translationally moving ones since the sample averaged cluster radius stays at its maximal value while $N_\Gamma$ decreases over time. For increasing particle mass the PFC force becomes less relevant compared to the alignment interaction meaning that the break-up of rotating clusters is delayed which expresses in the increasing decay times $\tau_d$ observed in fig.~\ref{fig2} for $v_0 = 0.35$.

In principle, the same mechanisms of formation and decay should apply to the case $v_0 = 0.4$. However, a higher active drive increases the acceleration along a peak's orientation of self-propulsion. Judging from the higher maximum $N_\Gamma$ values obtained and their rapid increase during the formation time scale, the activity suffices to considerably increase the probability of initiating a rotating cluster. As a consequence, we observe low $\tau_f$ times over the whole range of mass values. Increasing active drive also means lowering the relative strength of the PFC force leading to the stronger increase of $\tau_d$ for low mass values. The saturation of the decay time in the high mass regime may be explained with the high maximal area fraction close to $\eta_0$. The corresponding high number of rotating clusters makes destructive interactions between them inevitable thereby effectively limiting the decay time scale.

\section{Conclusion}
\label{sec:conclusion}

The distinction of translational and rotational clusters and the suggested formation and break-up mechanism for the latter qualitatively explain the found time scales of transient collective dynamics. The relaxation process to the steady traveling crystal state is delayed for increasing particle mass. However, we emphasize that this delay does not trivially scale with the time scale of inertial motion $\gamma^{-1}$ which only captures the relaxation of individual particle velocities. Instead, we find that the mass dependent formation and decay times of rotational cluster defects govern the time scale of the transient dynamics. Note that particle based simulations with interaction rules comparable to the ones used here also find transient circulating particle clusters before global collective motion emerges \cite{Grossman2008}.

So far we have neglected the convective term in the orientation dynamics of the active PFC model. We note that we also exemplary tested the case where additionally the convective term in the velocity dynamics is neglected and found no qualitative difference to the previous results. However, both convective terms in the underdamped model eqs.~\eqref{eq:model_for_ud_active_systems} originate from including inertia into the underlying microscopic equations of motion \cite{arold2019mean}. Therefore, they must in general, like the here discussed inertial time scale, be considered in underdamped active matter models. In the present active PFC model we may again include the convective term to the orientation dynamics. Then, the fully underdamped dynamics leads to the emergence of other states in the long time limit. In the parameter regime of the traveling crystal state we still observe domains of collectively moving density peaks with hexagonal symmetry. However, the orientation and velocity fields around peaks and thereby their associated propulsion mechanism change qualitatively. Furthermore, other domains of constant density are present where no flows are induced due to a vanishing orientation field. The density value is given from the temperature and cubic terms of the PFC functional in eqs.~\eqref{eq:functionals}. In \cite{arold2019mean} we have shown for a different system in more detail how inertially induced convective flows change the phase diagram relative to an overdamped description.

These examples demonstrate the relevance of our model for future works in the currently opening realm of underdamped active matter research \cite{Lowen2019Inertial}. Specifically, the emergence of motility induced phase separation (MIPS) in active systems has been recently predicted to depend on the inertial time scale of the single particles \cite{mandal2019motility}. With appropriately chosen interactions the underdamped model eqs.~\eqref{eq:model_for_ud_active_systems} or an alternation thereof might advance the topic further.

\section*{Acknowledgments}
Parts of this work have been supported by the German Research Foundation (DFG) by grant Schm 2657/4.

\section*{Author contribution statement}
MS conceptualized the project and guided the research. DA derived the model, carried out the simulations and analyzed the results. Both authors wrote the manuscript.

\bibliographystyle{epj}
\bibliography{references}
%
%
%

\end{document}